\documentclass[a4paper,11pt]{article}
\pdfoutput=1 

\usepackage{jheppub} 

\usepackage{graphicx}

\title{\boldmath Interacting Ghost Dark Energy in Complex Quintessence Theory}

\author[a]{Yang Liu}

\affiliation[a]{Faculty of Physics, Ludwig Maximillian Unisersity Munich,\\Theresienstrasse 37, D-80333 Munich, Germany}

\emailAdd{xijubear2020@Outlook.com}

\abstract{We employ a ghost model of interacting dark energy to obtain the equation of  state $\omega$ for  ghost  energy  density  in  an FRW  universe  in  complex  quintessence  theory.   We reconstruct the potential and study the dynamics of the scalar field that describes complex quintessence  cosmology.   We perform  $\omega-\omega'$ analysis and   stability  analysis  for  both non-interacting  and  interacting  cases   and  find  that  the same basic  conclusion as  for  the  real  model, where $\omega' = d\omega / d ln a$.  Taking account of  the  effect  of  the complex  part  and  assuming   the  real  part  of the quintessence  field  to be  a  “slow-rolling”  field,   we  conclude  that  the non-interacting  model cannot describe the real universe since this will lead to fractional energy density $\Omega_D > 1$, where $\Omega_D$ can be defined as the ratio of $\rho_D$  to $\rho_{cr}$. However, for the interacting case, if we take present $\Omega_D =0.73$, then we can determine that $b^2 = 0.0849$, where $b^2$ is the interaction coupling parameter between matter and dark energy. In the real quintessence model, $\Omega_D$ and $b^2$ are independent parameters,  whereas in the complex quintessence  model,  we conclude that there is a relationship between these two parameters.}

\begin{document} 
\maketitle
\flushbottom

\section{Introduction}
\label{sec:intro}
The accelerated expansion of the universe has been convincingly substantiated by observations on type Ia supernovae [1,2,3], Large Scale Structure [4,5] and Cosmic Microwave Background anisotropies [6]. In order to explain the reasons for the expansion of the universe, cosmologists have proposed many theoretical models. The most well-accepted explanation at present is dark energy. Dark energy has a negative pressure which leads to an accelerated expansion of universe. However, the nature and cosmological origin of dark energy has not yet been identified.\\ 
The most obvious candidate for  explaining the nature and origin of dark energy is the cosmological constant [7,8] which has the equation of state  $p=-\rho$.  However, this explanation itself inevitably leads to further difficulties, such as “the fine-tuning problem” etc. In view of this,  a series of alternative theoretical models have been proposed. In particular, a group of scalar field dark energy theories including quintessence [9], K-essence [10], tachyon [11], phantom [12], ghost condensate [13, 14] and quintom [15], braneworld models [16], interacting dark energy models [17], and Chaplygin gas models [18] etc. have been widely studied.\\
In addition to the degrees of freedom that already exist in standard cosmological models, most dark energy models introduce even further degrees of freedom. Introducing such further degrees of freedom requires studying their properties and the further consequences arising from them in modeling the universe. In view of this, those  new suggestions on the origin of dark energy have been made and which do not require introducing further unknown degrees of freedom  whilst yielding the necessary cosmic expansion with a dark energy value of the correct magnitude are of particular interest. Among these models, ghost dark energy (GDE) which uses the so-called Veneziano ghost to account for the observed accelerated expansion of the universe [19,20] is one well-known example.  In the GDE model, the cosmological constant is considered to originate  from the contribution of  ghost fields, which are predicted to exist according to low energy effective theory [21]-[25].  The ghosts are introduced in order to resolve the $U(1)$ problem [21]-[25]. It is claimed that the Veneziano ghost field provides  important physical  spacetime effects within non-topological or dynamic spacetime. The ghost field in curved space leads to a small vacuum energy density which is proportional to $\Lambda^3_{QCD} H$, where $\Lambda^3 $ is the QCD mass scale and $H$ is the Hubble constant [25]-[27]. The advantages of this model are that  it does not require the introduction of any new parameters or degrees of freedom and that it can be totally embedded in the standard model as well as general relativity. The dynamic behaviour of this model is presented in ref.[27]. \\
One initial example of scalar-field theory is so-called “quintessence” [9], which is described by a scalar field $Q$ having a slowly decreasing potential $V (Q)$. If the field evolves slowly enough, the kinetic energy density is less than the potential energy density, resulting in negative pressure, which means that the universe will experience an accelerated expansion. Zlatev, Wang and Steinhardt [28] as well as Huterer and Turner [29] have considered the basic properties of real ”quintessence” theory. In most of the subsequent papers, real scalar field cases were considered with only a few papers dealing with complex field cases. Gu and Hwang [30] pointed out the possibility of using a complex scalar field for the quintessence as a way of explaining the acceleration of the universe. Extending the ideas of Huterer and Turner [29] they derived reconstruction equations for the complex field, which demonstrated the feasibility of using complex scalar fields whilst maintaining the uniqueness of the inverse problem. However, Gu and Hwang [26] did not explore the possibility of using a dark energy model in a complex scalar field and more particularly in complex “quintessence” theory. In this paper, we consider the correspondence between ghost dark energy and the complex “quintessence” model with the latter being a source of the former. \\
We first review the basics of complex quintessence field theory as well as ghost dark matter and propose a correspondence between the ghost dark energy scenario and the complex quintessence model. We then reconstruct the potential and study the dynamics of a scalar field that describes complex quintessence cosmology.  In section 3, we preform the $\omega-\omega'$ analysis and the stability analysis for the non-interacting case, and we find that the basic conclusion is the same as the real model. Taking account of the effect of the complex part and assuming the real part of the quintessence  field  to be a  “slow-rolling”  field,  we suggest that a non-interacting model cannot describe the real universe since this will lead to $\Omega_D > 1$. In section 4, we conduct the $\omega-\omega'$ analysis and the stability analysis for the interacting case, which has the same result as for the real model. Considering the contribution of the complex part of quintessence,  regarding real part as a “slow-rolling” field and taking  $\Omega_D =0.73$, then we can determine that $b^2 = 0.0849$, which is consistent with the observational data.[31] In the real quintessence model, $\Omega_D$ and $b^2$ are independent parameters, whereas in the complex quintessence model these two parameters are interrelated.\\ 

\section{The Basics}
\subsection{The Complex Quintessence Field}
We consider the Friedmann-Robertson-Walker (FRW) metric
\begin{equation}\label{eq:2.1}
ds^2 = -dt^2 + a^2 (t) (\frac{dr^2}{1-kr^2} + r^2 d\Omega^2) 
\end{equation}
where $k$ is the curvature of space, and $k=0,1,-1$ for a flat, a closed and an open universe respectively. The action of the universe is given by
\begin{equation}\label{eq:2.2}
S = \int d^4 x \sqrt{-g} (\frac{1}{16\pi G} R + \rho_m + \mathcal{L}_\Phi)
\end{equation}
where $g$ is the determinant of the metric tensor $g_{\mu \nu}$, $G$ is the Newton’s constant, $R$ is the Ricci scalar, $\rho_m$ is the  density of ordinary matter, and $\mathcal{L}_\Phi$ is the Lagrangian density of the complex quintessence field $\Phi$ given by:
\begin{equation}\label{eq:2.3}
\mathcal{L}_\Phi = -\frac{1}{2} g^{\mu \nu} (\partial_{\mu} \Phi^{*}) (\partial_{\nu} \Phi) - V(|\Phi|)
\end{equation}
Here $\mu, \nu = 0,1,2,3$. We have assumed that in eq.$(2.3)$ that the potential $V$ depends solely on the absolute value of the complex quintessence field: $|\Phi|$.[30] \\
We can express $\Phi$ in terms of amplitude $\phi$ and  phase $\theta$ to as,
\begin{equation}\label{eq:2.4}
\Phi (x) = \phi(x) e^{i\theta(x) } 
\end{equation}
Or more precisely, $\Phi (t) = \phi(t) e^{i\theta(t) }$. We can then rewrite eq.$(2.3)$ as 
\begin{equation}\label{eq:2.5}
\mathcal{L}_\Phi = -\frac{1}{2} g^{\mu \nu} (\partial_{\mu} \phi) (\partial_{\nu} \phi) - \frac{1}{2} \phi^2 g^{\mu \nu} (\partial_{\mu} \theta) (\partial_{\nu} \theta) - V(\phi) 
\end{equation}
By employing the metric of eq.$(2.1)$, we have the following equations:
\begin{equation}\label{eq:2.6}
H^2 \equiv (\frac{\dot a}{a})^2 = \frac{8\pi G}{3} \rho- \frac{k}{a^2} =\frac{8\pi G}{3} \{\rho_m + \frac{1}{2} (\dot \phi^2+\phi^2 \dot \theta^2) + V(\phi)\}  - \frac{k}{a^2} 
\end{equation}
\begin{equation}\label{eq:2.7}
(\frac{\ddot a}{a})^2 = -\frac{4\pi G}{3} (\rho + 3p) = -\frac{8\pi G}{3} \{\frac{1}{2}\rho_m +  (\dot \phi^2+\phi^2 \dot \theta^2) - V(\phi)\}
\end{equation}
\begin{equation}\label{eq:2.8} 
\ddot \phi + 3H \dot \phi - \dot \theta^2 \phi + V'(\phi)  =0
\end{equation}
\begin{equation}\label{eq:2.9}
\ddot \theta + (2\frac{\dot \phi}{\phi} +3H) \dot \theta =0
\end{equation}
where $H$ is the Hubble parameter, dot and prime represent derivatives with respect to $t$ and $\phi$ respectively, $\rho$ is the energy density where $\rho=\rho_{\Phi} + \rho_m$, and $p$ is the pressure. Eqs. $(2.6)$–$(2.9)$ are the fundamental equations which govern the evolution of the universe. Eq. $(2.6)$ and $(2.7)$ are Friedmann equations for this model. From eq.$(2.5)$ we can derive the energy density $\rho_{\Phi}$ and pressure $p_{\Phi}$ to be:
\begin{equation}\label{eq:2.10}
\rho_{\Phi}= \frac{1}{2} (\dot \phi^2+\phi^2 \dot \theta^2) + V(\phi)
\end{equation}
\begin{equation}\label{eq:2.11}
p_{\Phi}= \frac{1}{2} (\dot \phi^2+\phi^2 \dot \theta^2) - V(\phi)
\end{equation}
Eq.$(2.9)$ can be solved to yield  the “angular velocity”, i.e. the time derivative of  $\theta$ which is given by 
\begin{equation}\label{eq:2.12}
\dot \theta = \frac{\omega}{a^3 \phi^2}
\end{equation}
where $\omega$ is an integration constant which is determined by the value of $\dot \theta$ initially or at some specific time. By using eq.$(2.12)$, eqs. $(2.6)$-$(2.9)$ can be rewritten in terms of $\phi$,  as presented in detail in ref.[30] to which reference is made.\\

\subsection{Ghost Dark Energy}
Following ref.[31], let us breifly review the ghost dark energy model briefly. For a non-flat FRW universe filled with dark energy and dust (dark matter), the corresponding Friedmann equation is written as
\begin{equation}\label{eq:2.13}
H^2 + \frac{k}{a^2} = \frac{8 \pi G}{3} (\rho_m + \rho_D) 
\end{equation} 
where $\rho_D$ and $\rho_m$ are the energy densities of dark energy and pressureless matter respectively.\\
The ghost energy density in standard cosmology is defined by
\begin{equation}\label{eq:2.14}
\rho_D = \alpha H 
\end{equation}
where $\alpha$ denotes a constant of order $\Lambda^3_{QCD}$ and $\Lambda_{QCD}$ represents the QCD mass scale and $H$ is the Hubble constant. Here $\Lambda_{QCD}$ is approximately equal to $100MeV$ and $H$ is approximately equal to $10^{-33}eV$, so $\Lambda^3_{QCD} H$ gives the correct order of magnitude $(3 \times 10^{-3} eV)^4$ for the presently observed dark energy value. This numerical coincidence is quite remarkable and also means that this model can obviate  the fine-tuning problem [19,20]. \\
We define the curvature energy density $\rho_k$ and the critical energy density $\rho_{cr}$ as usual as:
\begin{equation}\label{eq:2,15}
\rho_k = \frac{3k} {8\pi G a^2}
\end{equation}
\begin{equation}\label{eq:2.16}
\rho_{cr}= 3 M^2_p H^2
\end{equation}
We also introduce three fractional energy densities $\Omega_D$, $\Omega_m$ and $\Omega_k$:
\begin{equation}\label{eq:2.17}
\Omega_D = \frac{\rho_D} {\rho_{cr}} = \frac{\alpha} {3 M^2_p H}
\end{equation}
\begin{equation}\label{eq:2.18}
\Omega_m = \frac{\rho_m} {\rho_{cr}} = \frac{\rho_m} {3 M^2_p H^2}
\end{equation} 
\begin{equation}\label{eq:2.19}
\Omega_k = \frac{\rho_k} {\rho_{cr}} = \frac{k} {H^2 a^2}
\end{equation}
Then the Friedmann equation can then be written as:
\begin{equation}\label{eq:2.20}
1+ \Omega_k = \Omega_D + \Omega_m
\end{equation}
If there is no interaction between matter and ghost dark energy, i.e.,
\begin{equation}\label{eq:2.21}
\dot \rho_D + 3H \rho_D (1+\omega_D ) =0 
\end{equation}
\begin{equation}\label{eq:2.22}
\dot \rho_m + 3H \rho_m =0 
\end{equation}
then for the ghost dark energy density the equation of state [31,32] can be derived to be:
\begin{equation}\label{eq:2.23}
\omega_D = - \frac{1} {2-\Omega_D} (1- \frac{\Omega_k} {3} )
\end{equation}
If there is an interaction between matter and ghost dark energy, i.e.,
\begin{equation}\label{eq:2.24}
\dot \rho_D + 3H \rho_D (1+\omega_D ) = -Q 
\end{equation}
\begin{equation}\label{eq:2.25}
\dot \rho_m + 3H \rho_m = Q
\end{equation}
where $Q$ is the interaction term, then we define $Q$ as
\begin{equation}\label{eq:2.26}
Q= 3b^2 H (\rho_m +\rho_D) = 3b^2 H \rho_D (1+ u)
\end{equation}
with $b^2$ a coupling parameter. Here $u$ is defined as
\begin{equation}\label{eq:2.27}
u = \frac{\rho_m} {\rho_D} = \frac{\Omega_m} {\Omega_D} = \frac{1-\Omega_D} {\Omega_D}
\end{equation}
Under these assumptions for the ghost dark energy density,  the equation of state [31,32] can be derived to be:
\begin{equation}\label{eq:2.28}
\omega_D = - \frac{1} {2-\Omega_D} (1- \frac{\Omega_k} {3}+ \frac{2b^2} {\Omega_D} (1+ \Omega_k) )
\end{equation}
Further details of ghost dark energy can be found in ref. [31,32] to which reference is made.

\section{The Complex Quintessence Field as Ghost Dark Energy in an FRW Universe}
Considering first the case without interaction and establishing the correspondence between the energy density of the complex quintessence field and ghost dark energy, then using eq.$(2.10)$, $(2.12)$ and $(2.14)$, we obtain
\begin{equation}\label{eq:3.1}
\rho_D= \frac{1}{2} (\dot \phi^2+ \frac{\omega^2} {a^6 \phi^2}) + V(\phi) = \alpha H 
\end{equation}
For brevity, if we set
\begin{equation}\label{eq:3.2}
A= \frac{1}{2} (\dot \phi^2+ \frac{\omega^2} {a^6 \phi^2})  
\end{equation}
then we have 
\begin{equation}\label{eq:3.3}
V(\phi) = \alpha H - A  
\end{equation}
On the other hand, if we establish the correspondence between the energy density of the complex quintessence field and the ghost dark energy, then we have
\begin{equation}\label{eq:3.4}
\omega_{\Phi} \equiv \frac{p_{\Phi}}{\rho_{\Phi}} = \omega_D
\end{equation}
Combining eq.$(2.10)$, $(2.11)$ and $(2.23)$, we obtain
\begin{equation}\label{eq:3.5}
\frac{A - V(\phi)}{A + V(\phi)} =  - \frac{1} {2-\Omega_D} (1- \frac{\Omega_k} {3} )
\end{equation}
Rewriting the potential $V(\phi)$ gives
\begin{equation}\label{eq:3.6}
V(\phi) = \frac{3- \Omega_D - \frac{\Omega_k}{3}} {1- \Omega_D + \frac{\Omega_k}{3}} A
\end{equation}
Combining eq. $(3.3)$ and $(3.6)$, gives
\begin{equation}\label{eq:3.7}
\frac{3- \Omega_D - \frac{\Omega_k}{3}} {1- \Omega_D + \frac{\Omega_k}{3}} A = \alpha H - A 
\end{equation}
so that, then we have
\begin{equation}\label{eq:3.8}
H = \frac{1}{\alpha} (\frac{4-2\Omega_D}{1- \Omega_D +\frac{\Omega_k}{3}}) A
\end{equation}
In this article, we principally consider the case for $k=0$. We will conduct $\omega-\omega'$ analysis, stability analysis and further physical interpretation for the non-interacting model.\\
The $\omega-\omega'$ analysis is an important tool to discriminate different models as was first proposed in ref.[33] and subsequently  extended in ref.[34,35]. Here $\omega$ is the equation-of-state parameter $\omega \equiv p/\rho$, while $\omega'$ is the derivative of $\omega$ with respect to $ln a$. Based on ref.[31], in flat space $k=0$, so we have
\begin{equation}\label{eq:3.9}
\frac{d\Omega_D}{d ln a} = 3 \Omega_D \frac{1-\Omega_D}{2-\Omega_D}
\end{equation} 
and 
\begin{equation}\label{eq:3.10}
\omega_D = - \frac{1}{2-\Omega_D}
\end{equation} 
So that we obtain
\begin{equation}\label{eq:3.11}
\frac{d \omega_D}{d ln a} = -\frac{3 \Omega_D (1- \Omega_D)}{(2-\Omega_D)^3}
\end{equation}
Based on eq.$(3.10)$, eq.$(3.11)$ can be rewritten as
\begin{equation}\label{eq:3.12}
\frac{d \omega_D}{d ln a} = -3 \omega^3_D (2 + 1/\omega_D)(1+1/\omega_D)
\end{equation}
where $-1 < \omega_D <-\frac{1}{2}$. From eq.$(3.12)$, we know that $\frac{d \omega_D}{d ln a}$ is always negative, namely, $\omega$ decreases 
monotonically with the increase of $ln a$. Figure 1 shows the evolution trajectories of ${\omega-\omega'}$. From Fig.1, we know that when $\omega_D$ is about 0.8, $\omega_D'$ has a minimum. In addition, when $\omega_D$ equals to $-1$ or $-0.5$, $\omega_D'$ equals to 0.\\
\begin{figure}[htbp]
\centering
\includegraphics[scale=0.7]{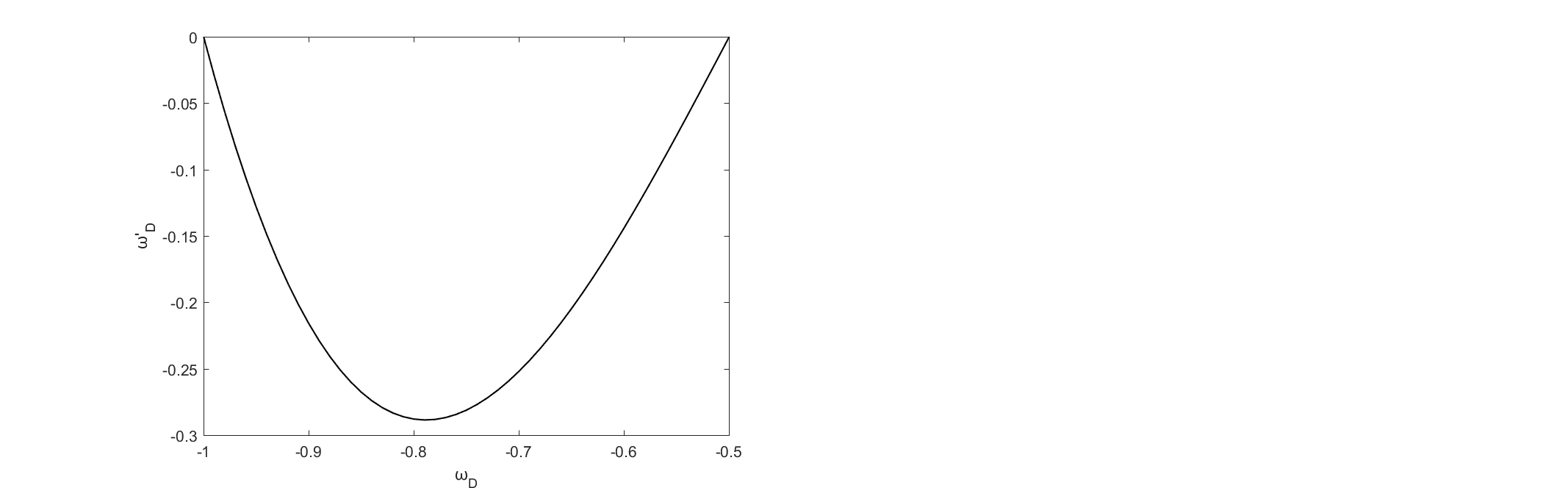}
\caption{Evolution trajectories of $\omega_D-\omega_D'$ for the non-interacting case}
\end{figure}\\
We can also determine the stability of this model by defining the velocity of sound  according to,
\begin{equation}\label{eq:3.13}
v^2_s (z) \equiv d p_D /d \rho_D = \frac{dp_D/dz}{d\rho_D/dz}
\end{equation}  
where $1+z = a^{-1}$ and $z$ is the redshift. In flat space, $k=0$, the Hubble constant $H$ can be written as [31]
\begin{equation}\label{eq:3.14}
H = \frac{\alpha (1+u)}{3 M^2_p}
\end{equation}  
where $M^2_p = 1/8\pi G$ and $u = \rho_m/\rho_D$. We can then obtain,
\begin{equation}\label{eq:3.15}
\frac{dH}{da} = -\frac{\alpha}{M^2_p \Omega_D a} \frac{1-\Omega_D}{2-\Omega_D}
\end{equation} 
Then we have
\begin{equation}\label{eq:3.16}
\frac{d\rho_D}{dz} = \alpha\frac{dH}{dz} =\alpha (-a^2) \frac{dH}{da} = \frac{\alpha^2 a}{M^2_p \Omega_D} \frac{1-\Omega_D}{2-\Omega_D}>0
\end{equation}
The stability of this model is therefore determined by the sign of $dp_D/dz$. Since $p_D = \omega_D \rho_D$, then
\begin{equation}\label{eq:3.17}
\frac{dp_D}{dz} = \frac{d\omega_D}{dz} \rho_D + \omega  \frac{d\rho_D}{dz} = -a \frac{d\omega_D}{dlna} \rho_D + \omega \frac{d\rho_D}{dz}
\end{equation}
Combining eq.$(3.11)$, $(3.13)$, $(3.16)$ and eq.$(3.17)$, we can then derive that 
\begin{equation}\label{eq:3.18}
v^2_s \equiv d p_D /d \rho_D = \frac{2\Omega_D - 2}{(2-\Omega_D)^2}
\end{equation}
If we require this model to be stable, then $v^2_s (z)$ must be larger than zero. Recalling eq.$(3.18)$ this means that $\Omega_D > 1$ which is impossible. Figure 2 shows the evolution of $v^2_s$ against $\Omega_D$ for the non-interacting case. Since the magnitude of the speed of sound cannot be negative , a non-interacting ghost dark energy dominated universe in the future cannot be expected to be the fate of the universe. In fact, this is a general conclusion for  ghost dark energy model irrespective of whether the complex part of scalar field is considered or not. Similar results are presented in  ref.[32] to which reference is made.\\
\begin{figure}[htbp]
	\centering
	\includegraphics[scale=0.7]{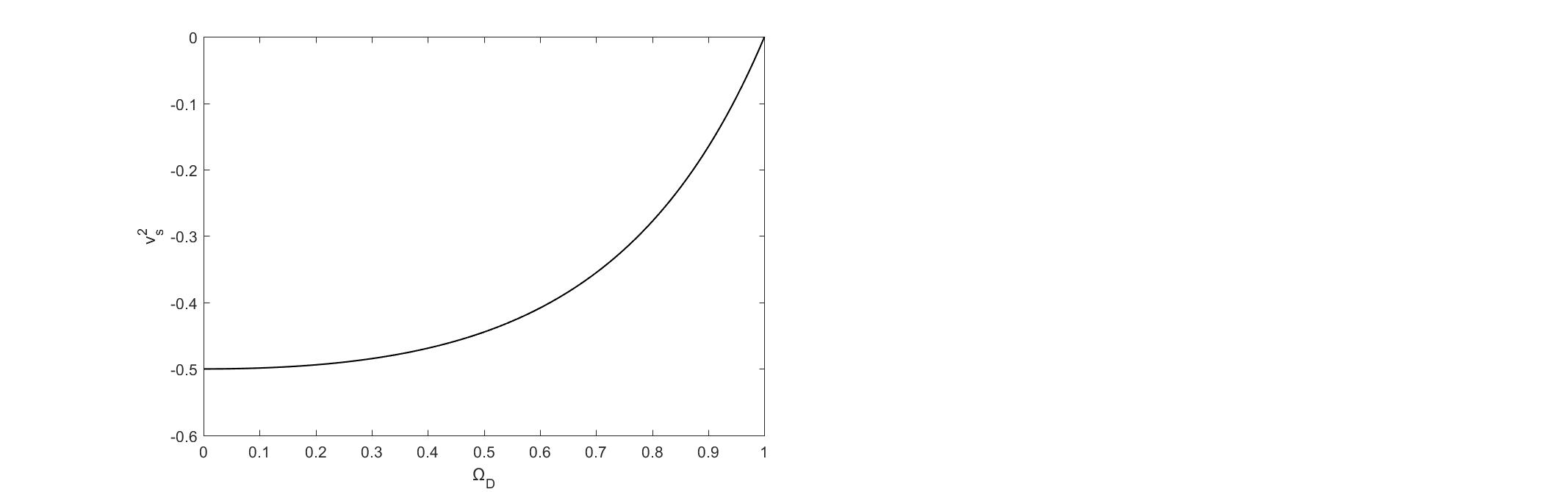}
	\caption{Evolution of the speed of sound squared $v^2_s$ versus $\Omega_D$ for the non-interacting case}
\end{figure}\\  
We can expect that the above $\omega-\omega'$ analysis and stability analysis for the complex quintessence model to yield the same results  when using  the real quintessence model, since the two analyses  only  involve  the  ghost  dark  energy  model  which is entirely unrelated to  any   effect  from the complex  part.  We will now consider  the  effect  of  the  complex  part  of  the quintessence  field.   In  this article, we restrict our  considerations to the effect of a “slow-rolling field”. \\
Combining eq.$(2.10)$, $(2.11)$ and $(3.2)$, we can derive,
\begin{equation}\label{eq:3.19}
\alpha H \frac{1-\Omega_D}{2-\Omega_D} = \dot{\phi}^2 + \frac{\omega^2}{a^6 \phi^2}
\end{equation}
Considering only a “slow-rolling” field, i.e., ignoring the term $\dot{\phi}^2$, we have
\begin{equation}\label{eq:3.20}
\alpha H \frac{1-\Omega_D}{2-\Omega_D} = \frac{\omega^2}{a^6 \phi^2}
\end{equation}
So that
\begin{equation}\label{eq:3.21}
\phi = \frac{\omega}{a^3} \sqrt{\frac{2-\Omega_D}{(1-\Omega_D)\alpha H}}
\end{equation}
We consider only the positive solution. Since $\dot{\phi} = H \frac{d \phi}{d ln a}$ and $H$ cannot be ignored, then $\frac{d \phi}{d ln a} \approx 0$. Combining eq.$(3.9)$, $(3.15)$ and $(3.21)$, we have
\begin{equation}\label{eq:3.22}
\frac{d \phi}{d lna} = \frac{-3\omega}{a^3} \sqrt{\frac{2-\Omega_D}{(1-\Omega_D)\alpha H}} + \frac{\omega}{2 a^3} \frac{1}{\sqrt{\frac{2-\Omega_D}{(1-\Omega_D)\alpha H}}} \frac{3(\Omega^2_D - 2\Omega_D +2)}{\alpha H (\Omega^2_D - 3\Omega_D +2)} \approx 0
\end{equation}
After rearranging terms, we obtain,
\begin{equation}\label{eq:3.23}
\Omega^2_D - 6\Omega_D +6 = 0
\end{equation}
Whose solutions are  $\Omega_D = 3 \pm \sqrt{3}$, which is inconsistent with the assumption  $\Omega_D < 1$. We can therefore conclude that, if we regard the complex quintessence field as ghost dark energy and make the “slow-rolling” field assumption, the non-interacting case cannot be used to describe the real universe. We therefore need to consider the case of interaction between matter and ghost dark energy. In the next section, we will determine the value of $b^2$ in eq.$(2.26)$.

\section{Interacting Ghost Dark Energy in an FRW Universe in Complex Quint-essence Theory}
We now  consider the case of interaction between matter and ghost dark energy. Combining eq.$(2.17)$, $(2.19)$ and $(2.28)$, we obtain
\begin{equation}\label{eq:4.1}
\omega_D=-\frac{1} {2-\frac{\alpha} {3 M^2_p H} } \{1-\frac{k} {3 H^2 a^2} + \frac{6 b^2 M^2_p H} {\alpha} (1+\frac{k} { H^2 a^2} ) \}  
\end{equation}
In particular, for a flat universe, $k=0$ and $\Omega_k = 0$, then
\begin{equation}\label{eq:4.2}
\omega_D=- \frac{1} {2-\Omega_D} (1+ \frac{2b^2} {\Omega_D} )=-\frac{1} {2-\frac{\alpha} {3 M^2_p H} } (1 + \frac{6 b^2 M^2_p H} {\alpha})   
\end{equation}
For brevity, we set 
\begin{equation}\label{eq:4.3}
B=-\omega_D= \frac{1} {2-\frac{\alpha} {3 M^2_p H} } \{1-\frac{k} {3 H^2 a^2} + \frac{6 b^2 M^2_p H} {\alpha} (1+\frac{k} { H^2 a^2} ) \}  
\end{equation}
If we establish the correspondence between the energy density of the complex quintessence field and the ghost dark energy, then we obtain,
\begin{equation}\label{eq:4.4}
\omega_D= \omega_{\Phi}  
\end{equation}
that is to say,
\begin{equation}\label{eq:4.5}
-B= \frac{A - V(\phi)}{A + V(\phi)}   
\end{equation}
so that we can obtain
\begin{equation}\label{eq:4,6}
V(\phi)= -\frac{B + 1}{B -1} A   
\end{equation}
Combining eq.$(3.3)$ and $(4.6)$, we have
\begin{equation}\label{eq:4.7}
-\frac{B + 1}{B -1} A = \alpha H - A    
\end{equation}
therefore
\begin{equation}\label{eq:4.8}
H = -\frac{2}{\alpha} \frac{1}{B -1} A = \frac{2}{\alpha} \frac{1}{\omega_D+1} A    
\end{equation}
In order to be self-consistent, $B$ must be smaller than $1$, i.e.,
\begin{equation}\label{eq:4.9}
\frac{1} {2-\Omega_D} \{1-\frac{k} {3 H^2 a^2} + \frac{2 b^2 } {\Omega_D} (1+\frac{k} { H^2 a^2} ) \} < 1    
\end{equation}
To ensure the required accelerated expansion of the universe, based on eq.$(2.7)$ we have,
\begin{equation}\label{eq:4.10}
\rho_m < 2(V(\phi)-(\dot{\phi}^2 + \phi^2 \dot{\theta}^2 )) = 2 V(\phi) - 4A 
\end{equation}
Considering eq.$(3.3)$, we have,
\begin{equation}\label{eq:4.11}
\rho_m < 2 V(\phi) - 4A = 6 V(\phi) - 4\alpha H
\end{equation}
therefore
\begin{equation}\label{eq:4.12}
V(\phi) > \frac{2}{3} \alpha H 
\end{equation}
Combining eq.$(3.3)$ and $(4.12)$, we have as the constraint for $V (\phi)$, 
\begin{equation}\label{eq:4.13}
 \frac{2}{3} \alpha H < V(\phi) < \alpha H
\end{equation}
We will now perform the $\omega-\omega'$ analysis for interacting case in flat space $k=0$. Considering eq.$(2.28)$, we have
\begin{equation}\label{eq:4.14}
\omega_D = -\frac{1}{2-\Omega_D} (1+ \frac{2b^2}{\Omega_D})   
\end{equation}
So that
\begin{equation}\label{eq:4.15}
\omega_D' = \frac{\frac{2b^2}{\Omega_D^2} \Omega_D' (2-\Omega_D) - \Omega_D'(1+ \frac{2b^2}{\Omega_D})}{(2-\Omega_D)^2}  
\end{equation}
where $'$ denotes the derivative with respect to $lna$. Based on ref.[31], we have
\begin{equation}\label{eq:4.16}
\frac{d\Omega_D}{d lna} = \frac{3}{2} [1- \frac{\Omega_D}{2-\Omega_D} (1+ \frac{2b^2}{\Omega_D})]    
\end{equation} 
Therefore,
\begin{equation}\label{eq:4.17}
\omega_D' = \frac{[1-\frac{\Omega_D}{2-\Omega_D}(1+\frac{2b^2}{\Omega_D})](\frac{6b^2}{\Omega_D} -\frac{3}{2} \Omega_D - 6b^2)}{(2-\Omega_D)^2}  
\end{equation}
In particular, when $b^2=0$, eq.$(4.13)$ becomes eq.$(3.12)$. Figure 3 shows the evolution of $\omega-\omega'$ for different values of $b^2$. Here we have taken $\Omega_D = 0.73$.[31] From figure 3, we conclude that for the interacting case, there exists a “overlapping” region in which a single value of $\omega_D$ corresponds to two possible values of $\omega_D'$. In particular, the width of the region becomes narrower with the increasing of $b^2$. In contrast to this,  for  the  non-interacting  case,  there is no  “overlapping”  region  at all. This result is different from that in an agegraphic dark energy model.[36] In fact, the $\omega-\omega'$ analysis is a useful dynamic analysis for discriminating different dark energy models.[36]\\
\begin{figure}[htbp]
	\centering
	\includegraphics[scale=0.7]{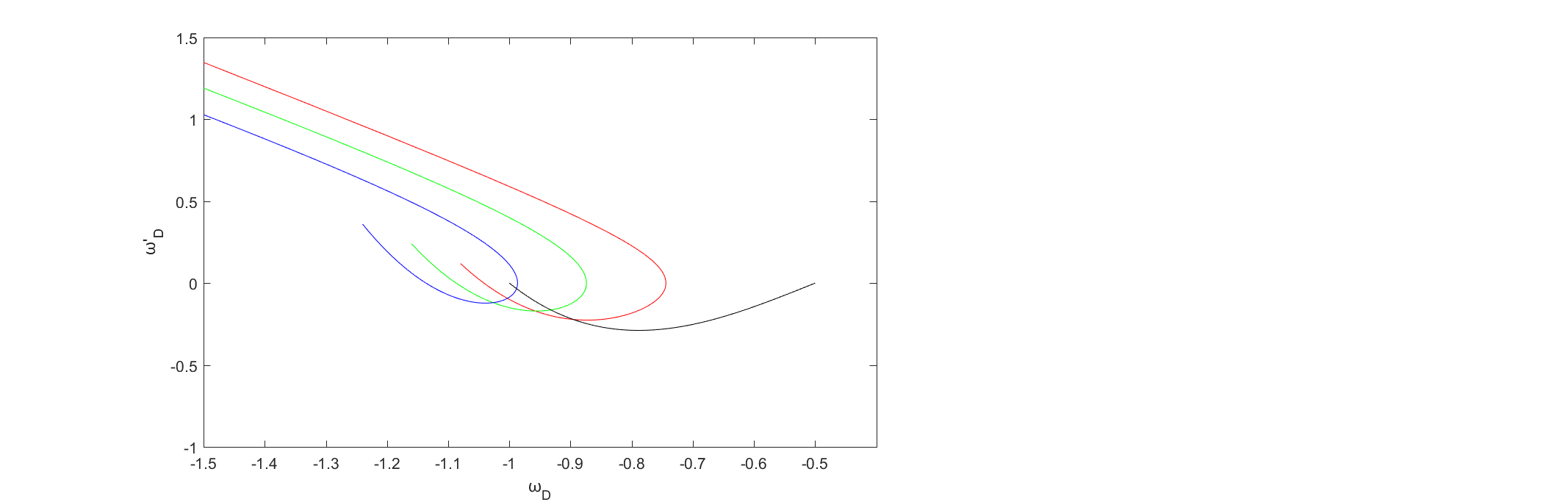}
	\caption{Evolution trajectories of $\omega_D-\omega_D'$ for different values of $b^2$. The red line corresponds to $b^2 = 0.04$, the green line to $b^2=0.08$, the blue line to $b^2 =0.12$ and the black line to $b^2 =0$.}
\end{figure}\\
Continuing with the above,  we can also determine the stability of this model for the interacting case. Using a  similar calculation, we obtain,
\begin{equation}\label{eq:4.18}
\frac{dH}{da} = -\frac{\alpha}{2a \Omega_D M^2_p} [1- \frac{\Omega_D}{2-\Omega_D} (1+ \frac{2b^2}{\Omega_D})] 
\end{equation}
\begin{equation}\label{eq:4.19}
\frac{d\rho_D}{da} = \frac{\alpha^2 a}{2 \Omega_D M^2_p} [1- \frac{\Omega_D}{2-\Omega_D} (1+ \frac{2b^2}{\Omega_D})] 
\end{equation}
and 
\begin{equation}\label{eq:4.20}
\begin{aligned}
\frac{dp_D}{da} = &-\frac{1}{(2-\Omega_D)^2} \frac{3}{2} \Omega_D a \rho_D (\frac{4b^2}{\Omega_D^2} - \frac{4b^2}{\Omega_D} - 1) [1- \frac{\Omega_D}{2-\Omega_D} (1+ \frac{2b^2}{\Omega_D})] \\
& - \frac{1}{2-\Omega_D} (1+ \frac{2b^2}{\Omega_D}) \frac{\alpha^2 a}{2 \Omega_D M^2_p} [1- \frac{\Omega_D}{2-\Omega_D} (1+ \frac{2b^2}{\Omega_D})] 
\end{aligned}
\end{equation}
The square of the velocity of  sound can then be obtained as,
\begin{equation}\label{eq:4.21}
v^2_s \equiv d p_D /d \rho_D = \frac{2\Omega^2_D-2\Omega_D+6b^2 \Omega_D-8b^2}{\Omega_D (2-\Omega_D)^2}
\end{equation}
For the model to be stable, then $v^2_s$ must be larger than zero. This means that $\Omega_D$ is larger than $\frac{4}{3}$, which is impossible. We can therefore conclude that for the interacting case, the model is still unstable and cannot conclude that a ghost dark energy dominated universe will be the future is the fate of the real universe. Figure 4 shows the evolution of $v^2_s$ against $\Omega_D$ for different $b^2$. Further details of the stability analysis can be found in ref.[32].\\
\begin{figure}[htbp]
	\centering
	\includegraphics[scale=0.7]{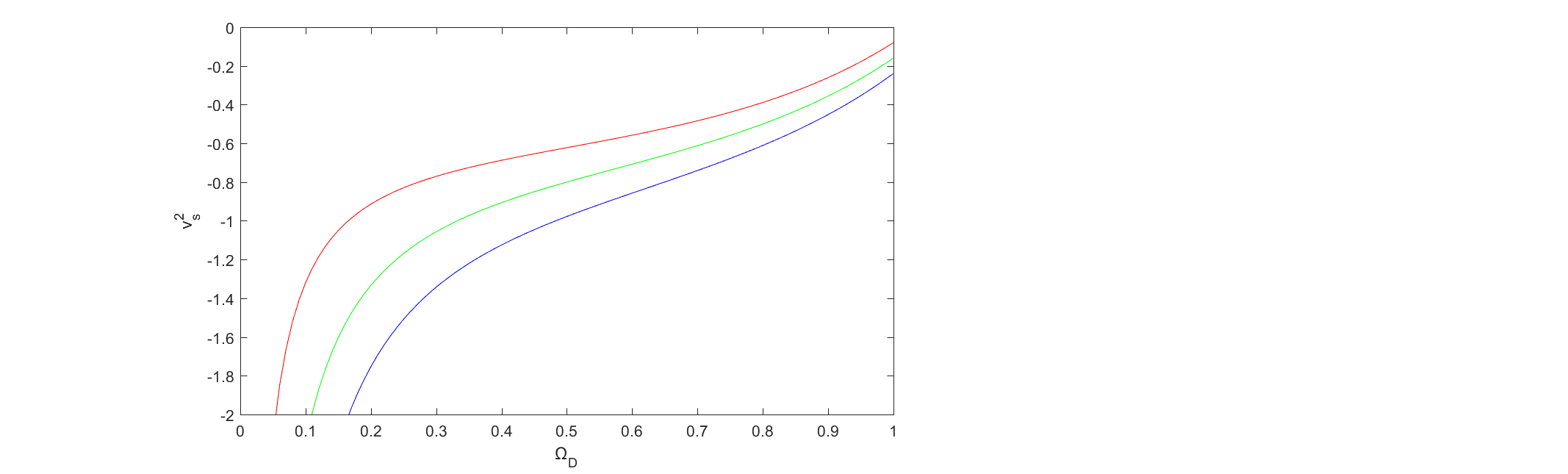}
	\caption{Evolution of the  speed of sound squared $v^2_s$ versus $\Omega_D$ for different values of $b^2$. The red line corresponds to $b^2 = 0.04$, the green line to $b^2=0.08$ and the blue line to $b^2 =0.12$. }
\end{figure} \\
The above $\omega-\omega'$ analysis and stability analysis yield the same results with a real quintessence scalar field, since the two analyses only involve the ghost dark energy model and do include the effect of the complex part of quintessence.  We will now consider the effect of the complex part  of  the quintessence  field  for the interacting  case.  As  mentioned  above,  we  only consider the effect of a “slow-rolling” field.\\
As  for the interacting case in flat space, we can then obtain
\begin{equation}\label{eq:4.22}
\alpha H \frac{\Omega_D (1-\Omega_D)-2b^2}{\Omega_D (2-\Omega_D)} \approx  \frac{\omega^2}{a^6 \phi^2}
\end{equation}
Therefore, we have
\begin{equation}\label{eq:4.23}
\phi = \frac{\omega}{a^3} \sqrt{\frac{\Omega_D (2-\Omega_D)}{\Omega_D (1-\Omega_D)-2b^2} \frac{1}{\alpha H}}
\end{equation}
If we require $\frac{d\phi}{dlna} \approx 0$, then we obtain
\begin{equation}\label{eq:4.24}
\begin{aligned}
0 \approx & -6 \Omega_D (2-\Omega_D) [\Omega_D (1-\Omega_D) -2b^2]\\
& +[3- \frac{3\Omega_D}{2-\Omega_D} (1+\frac{2b^2}{\Omega_D}) - 3\Omega_D + \frac{3\Omega_D^2}{2-\Omega_D}(1+\frac{2b^2}{\Omega_D})] [\Omega_D(1-\Omega_D)-2b^2] \Omega_D\\
& -(2-\Omega_D) [\frac{3}{2}\Omega_D -\frac{3}{2}\frac{\Omega_D^2}{2-\Omega_D}(1+\frac{2b^2}{\Omega_D}) - 3\Omega_D^2 + \frac{3\Omega_D^3}{2-\Omega_D}(1+\frac{2b^2}{\Omega_D})] \\
& + \frac{3}{2}\Omega_D [\Omega_D (1-\Omega_D) - 2b^2][2-\Omega_D - \Omega_D(1+\frac{2b^2}{\Omega_D})] \\
\end{aligned}
\end{equation}
If we take the present $\Omega_D = 0.73$ [31] and insert this value in eq.$(4.24)$, we have 
\begin{equation}\label{eq:4.25}
6.2424 b^4 + 8.0893 b^2 - 0.7317 =0
\end{equation}
from which it follows that the only reasonable solution for $b^2$, is $b^2 = 0.0849$. This numerical result for $b^2$ is consistent with the observational result of $b^2$ [31]. In ref.[31], the observational result is $b^2 = 0.09$. However, in the previous research, people only considered the real part of quintessence field so that $b^2$ was a free parameter in this model and could only be determined by observation. In this article, we consider the effect of the complex part of the quintessence field. We find that the non-interacting case cannot describe the real universe and an interaction between matter and dark energy must be added. In fact, eq.$(4.24)$ defines  the relationship between $\Omega_D$ and $b^2$.
Therefore, $\Omega_D$ cannot be an arbitrary value between $0$ and $1$. Its value must instead ensure the existence of a positive solution of eq.$(4.24)$ for $b^2$. \\

\section{Summary and Discussion} 
To address the problem of accounting for the accelerated expansion of the universe and due to an absence of knowledge in this domain, theoretical cosmologists have considered a variety of dark matter candidates to explain this phenomenon. The ghost dark matter (GDE) model which was proposed by F.R. Urban and A. R. Zhitnitsky [19] as well as N. Ohta [20] has been studied in several papers [27,31,32]. Without introducing further unknown degrees of freedom, this model predicts  cosmic expansion based on dark energy with a  correct magnitude.\\
Most previous papers focused on the effects of dark energy using different real scalar field theories, such as quintessence [9], K-essence [10], tachyon [11] etc. However, only a few studies have investigated the possibility of using an approach based on complex scalar fields. Gu and Hwang [30] pointed out the possibility of using complex scalar fields for quintessence as a way of explaining the acceleration of the universe. Complex scalar fields are already known to play an important role in elementary particle physics such as in the mass generation mechanism (Higgs mechanism). The physics of  complex fields therefore deserves attention as a way of constructing practical models of our universe.\\
In this paper, we have used the ghost model of interacting dark energy to obtain the equation of state for the ghost energy density in an FRW universe in complex quintessence theory. We  have derived  the  potential  and  studied  the  dynamics  of  the  scalar  field  that  describe complex quintessence cosmology.  In section 3, we performed the $\omega-\omega'$ analysis and the stability analysis  for  the  non-interacting  case   and found  that  the  basic  conclusion  is  the  same as  for the  real  model. Taking into account   the  effect  of  the complex  part  and  regarding  the  real  part  of the quintessence  field  as  a  “slow-rolling”  field,  we  concluded  that  the non-interacting  model  cannot describe the real universe since this will lead to $\Omega_D > 1$. In section 4, we performed the $\omega-\omega'$ analysis and the stability analysis for the interacting case, which lead to the same result as for the real model.  Considering  the  contribution  of  the complex  part  of  quintessence, regarding the real  part  as  a  “slow-rolling”  field and taking the value of  $\Omega_D =0.73$ we determined the value of $b^2 = 0.0849$, which is consistent with the observational data value $b^2=0.09$.[31] In the real model, $\Omega_D$ and $b^2$ are independent parameters, whereas  in  the  complex  model,  we  found an interrelationship between these two parameters.\\
Future work can be directed  along at least four lines  of further research. Firstly, we can establish the correspondence between the complex quintessence field and other dark energy scenarios. In fact, since the ghost dark energy model is unstable, a future universe dominated by ghost dark energy cannot be the fate of the real universe. In subsequent work, we intend to correlate the complex quintessence model and  the  holographic dark energy model as well as compare these results with the ghost dark energy model. Secondly, we can generalize real scalar-tensor field theories, such as the Brans-Dicke theory to complex versions and then consider dark energy models in generalized scalar-tensor field theory. Thirdly, we can combine complex scalar field theory with elementary particle physics theory to help us build practical models of the universe. Finally, we need to consider how to ensure compatibility of  these results with fundamental theories, such as string theory and loop quantum gravity. There is therefore great potential for development of this work in the future.\\





\end{document}